# A CLOSED NON-COLLAPSING 3-D UNIVERSE PREDICTING A NEW SOURCE OF GRAVITY AND DARK MASS

## Charles B. Leffert

Wayne State University, Detroit, MI 48202
(C_Leffert@wayne.edu)

**Abstract.** Could one start from scratch, ignore relativity theory and quantum theory, create and expand our 3-D universe with no singularities, have the mathematical model predict correctly all of the cosmological parameters, provide the origins and understanding of time, energy, gravity and some of quantum behavior inside our 3-D universe and do all of that with just high school algebra and a little calculus?  The answer is: "Yes, it has been done and includes many falsifiable predictions[1,2]."  For the mathematical model, some new physical concepts are needed together with some older physical concepts such as the scaling of radiation and matter with the expansion and the idea that clumps of mass locally curve our 3-D space.  Instead of assuming "roll-up"-dimensional changes of space that occurred sometime in the past, here the major postulate is that such an ongoing dimensional change of space drives our universe yet today.  The first paper [3], astro-ph/0102071, presented the predicted magnitudes of the new model for supernova Ia and the excellent fit with the astronomical measurements without the cosmological constant. This second paper will present more of the theory and other important predictions.

## PROBLEMS WITH CURRENT PHYSICS

A number of fundamental problems hint that some of our basic physical assumptions are not correct.  There is much motion in our 3-D universe but no recognition of a "power plant" that supports that activity.  Could it be that there is a fundamental dynamic at work in our universe that we have not yet recognized and accounted for?

It is well recognized that our two major theories, relativity and quantum theory, are not compatible.  Limited to three spatial dimensions, relativity theory predicts an infinite-density singularity inside black holes.  The big-bang cosmological model, based on the general theory of relativity, begins with a singularity and allows another possible infinite-density singularity in the collapse of a closed universe.  The predicted vacuum energy density of particle physics is inconsistent by 120 orders of magnitude.

Our present physics rests upon the fundamental concepts of space, time and energy.  To the question: "What <u>is</u> space?" -- or time?" -- or energy?" there are no deeper fundamental concepts to provide the answers.  Time is just a symmetric mathematical parameter added to physics from the outside. There is no conservation law for space. Empty space, or the "vacuum," is said to just "expand" but now it has been given added attributes such as energy – but no one knows what "energy" is.[2] Finally, there is an even deeper theoretical problem of the role played by the mathematical concept of *continuum* in physics.[4,5]

## OVERVIEW

It is proposed that our closed universe of three spatial dimensions is the surface of an expanding four-dimensional ball.  It expands not because of an initial big-bang explosion of all the enormous mass of our universe but because of a process, called "spatial condensation," in which Planck-size four-dimensional (4-D) spatial cells are produced. This process began in a symmetry-breaking event in an embedding m-dimensional Euclidean epi-space. This event led to the compaction of a small 4-D ball and our 3-D universe on its surface.

Immediately the contrast is evident with the big-bang model for a closed universe in which all of the mass-energy of our universe explodes from a point and expands as a 3-sphere and the so-called "attractive" gravitational force of its contents slows the expansion to a halt and then accelerates its collapse back to a point.

With the new idea of the condensation of a higher dimensional space to produce an ever-expanding 4-D ball, gravity has no effect whatsoever on the expansion rate and can only form dimples in the surface of the 4-D core. So a new theory of gravity must be derived from the new idea of spatial condensation. These dimples in the 4-D ball must account for what we have heretofore considered an "attractive force" and the geometry of those dimples must predict the same 3-D curvature of 4-D space as that of the predicted "hyperspace" of Einstein's relativity theory. Most important, the new model must predict a new concept of black holes where general relativity predicts a singularity of infinite mass density inside the black hole.

Quantum theory is concerned with the behavior of radiation and matter inside our 3-D universe. But radiation and matter cannot come into existence until our 3-D universe has been created. Thus the creation (cosmogony) and following expansion (cosmology) of the 4-D ball (or "4-D core") has to do with non-3-D spatial entities independent of resulting processes inside our 3-D universe. Indeed, the gravitational constant G, the speed of light C, and Planck's constant h must not appear explicitly in either the mathematical model of creation or that of the expansion of our universe. Those fundamental constants must appear when we begin to account for our measurements of the radiation, matter and gravity inside our 3-D universe.

One can begin to develop a predictive model based on this SC-idea as follows. Planck's natural system of units suggests a 4-D hypercube (call it a "4-D planckton" or "4-D pk") for the fundamental unit cell of the growing 4-D ball with spatial condensation reproduction every Planck unit of time, but only on the surface of those existing 4-D planckton exposed to the embedding epi-space, and during expansion only on the surface of the 4-D core.

Somehow what we call time must be related to the total number of 4-D planckton available for continued spatial condensation. A workable mathematical definition of a new *cosmic time* will be the key to a possible successful development of a new model for the expansion of our universe.

Present cosmological models relate the scale factor R to the assumed parametric time via differential equations with subsequent integration to determine the relation between time and the scale factor R. The new SC-model must begin (and does) instead with an explicit definition of cosmic time, $t=t(R)$, where the scale factor R is the radius of the 4-D core. Differentiation of $t(R)$ will then lead to all of the pertinent cosmological parameters. Much more is demanded. The new cosmic time must be asymmetric as is the spatial condensation process and as is our subjective notion of time and it must increase monotonically from the first 4-D pk produced on into the distant future.

Instead of an independent variable as in present physics, cosmic time becomes a dependent variable -- $dt/dN_4$ is the rate of production of cosmic time per new 4-D pk produced. Cosmic time or age $t(R)$ is thus a summation of the past <u>resistance</u> to spatial condensation. The defining equation for cosmic time must account for the history of changing resistance (different scaling of density with R) to the SC-process as our 3-D universe passes through the eras of radiation, matter and final dark-mass domination.

In the SC-model dark mass is not matter. Indeed, it is not even "3-D stuff" and does not scale with the expansion as does matter. The creation begins with the production of the first 4-D planckton of dark mass and thus interacts in the expansion only gravitationally with matter and radiation. As was seen in the first paper,[3] it was the new cosmic time and the new scaling of dark mass with the expansion that accounted for the excellent fit with the supernova Ia data with no cosmological constant.

The smaller spatial building blocks from the mother epi-universe that condense to form one 4-D pk are called m-D pk (m~10). These m-d pk arrive by the columns to all massive condensation sites in our 3-D universe. They also arrive with an impact that curves the 3-D surface and forms the dimple in the 4-D core. We must derive the new gravity from those impacting m-D pk and the curvature of that dimple. That new source of gravity must also be powerful enough to predict the 4-D geometry of black holes.

The stage for further development has now been set for *space* and it has been set for *time* but what about *energy*? Feynman wrote, "… we have no knowledge of what energy *is*."[6] Spatial condensation must also account for *energy*. But there are many forms of energy: mass-energy, kinetic energy, potential energy, vacuum energy, nuclear energy, etc. but very few adjustable concepts to do the accounting.

To start, *energy* is defined as: "*the rate of spatial condensation, pk s$^{-1}$ or pks*", but immediately, to account for the rest mass characteristic of energy, the definition is qualified to: *energy* is defined as: "*the rate of spatial condensation by <u>persistent</u> columns of arriving m-D pk.*" To account for kinetic energy, it is noted that those columns of arriving m-D pk can be at an angle with respect to the radius of the 4-D core as sketched in Fig. 2-1, and that angle changes with velocity or momentum and probably with epi-resistance – so at the same time we may account for a new source of inertia – but this accounting is at a cost.

That cost is fundamental and its price is high because it says that there is a preferred reference frame, namely the comoving frame in our 3-D universe, and that conclusion contradicts the very principle of relativity. Scientists have not found any evidence for a preferred reference frame, apart from the CBR, in their local physics. So the saving grace may be that, in principle, it becomes measurable only on cosmic scales.

Only with a predictive mathematical model can we begin to sort out the truth from fiction of the above postulates. Astronomical measurement is the final judge.

## THE MODEL

**The Beginning.** After many failed attempts, the "beginning" was the successful "spatial condensation" production in the higher-dimensional m-D epi-universe of the first lower-dimensional, but larger, 4-D spatial cell – a hypercube of Planck size $10^{-33}$ cm. This foreign object acted as a catalytic site for the production of yet another 4-D unit cell and then they both reproduced and an exponential production was underway. Each reproduction occurred every Planck time of $10^{-43}$ s. After $10^{-33}$ s, all $10^{139}$ of these 4-D spatial cells (call them 4-D "planckton" to be abbreviated "4-D pk") were driven into a stable but growing 4-D ball (also called a "4-D core") (radius R~72 cm) with our 3-D universe as its expanding surface. The creation model is described in detail elsewhere, [1,2] so only those very significant concepts that effect the expansion model and 3-D physics will be described here. The creation model requires only one input parameter.

Two types of 4-D pk are produced during this period of creation: (1) a core acceptable "c-type pk" and (2) a core rejectable "x-type pk". The c-type pk produce only c-type pk. The x-type pk also produce c-type pk but randomly, on average, one of four reproductions is another x-type pk.

At the end of the 4-D core formation, spatial condensation continues but only on the surface of the 4-D core. The internal c-type 4-D pk are cut off from further reproduction. The outer layer of c-type 4-D pk are in violent agitation from the spatial condensation but become quickly bonded to the 4-D core because that outer surface begins with a radial expansion rate of $dR/dt \sim 10^{24}$ times the speed of light. The major source of spatial condensation for the expansion of the 4-D core and our 3-D universe would seem to be the condensation on the surface layer apart from the presence of radiation or matter. That is, by definition, the "vacuum" appears to manifest "energy" as a rate of spatial condensation but, as we will see, the model predicts that the "vacuum energy" cannot be measured.

The minor $10^{34}$ rejected x-type 4-D pk continue to produce both c-type and x-type pk. We recognize the rejected x-type pk as dark mass because they dimple the 4-D core and thus "gravitate" and since they are not even 3-D stuff, dark mass does not otherwise interact with either radiation or matter. Our 3-D radiation and matter also make excellent condensation sites, producing dimpled curvature and responding to other dimpled curvature.

Such a story of creation, seemingly pulled out of the "blue", is worthy of study only if it provides further fundamental understanding of our universe. Again, note that this creation and beginning of expansion does not conflict with quantum theory since spatial condensation does not involve the quantum particles of radiation and matter. On the other hand spatial condensation may provide the missing dynamic for the source of quantum behavior. Already we can begin to account for non-local quantum behavior by the above-implied speed of communication through the epi-universe at a rate $C^+$ that must be greater than $10^{24}$ C. "Sum-over-histories" may actually occur in epi-space.

The impacting m-D pk from epi-space to the "massive" condensation sites dimples the surface of the 4-D core and thus curves 3-D space. So a new model of gravity must be derived when we get to the physics inside our 3-D universe. The spatial "essence" of radiation and matter has not yet been accounted for but a reproducing dark mass has been introduced as x-type 4-D pk. There is no such entity in current theory so a new scaling law must be introduced for it in the expansion model and that scaling law must make dark mass become the present and future dominant mass in our universe. Since it reproduces in place, that means it must occur only in growing clumps in our 3-D universe which will have an enormous influence on how the structure of matter evolves in our universe.

To proceed further in defining the initial conditions for the expansion, more details are needed. As the two types of free 4-D pk were driven into the geometry of the 4-D ball, smaller "corelets" formed bigger corelets until, like our moon, much of the final surface features (craters) were fixed by the last smaller objects to impact the final large object.

As the corelets formed, the x-type pk always remained on the surface of the corelet. But c-type pk only produce c-type pk, so many of the last impacting corelets could contain few, if any, x-type pk. When such a corelet merged with the 4-D core it would sweep a 3-D volume clean of any already present x-type 4-D pk and, again like the moon, leave the x-type pk ejecta rather smoothly concentrated around the periphery of the miniature void as seeds for future production of galaxies and clusters of galaxies.

Somehow the stable 3-D spatial makeup of matter (and anti-matter) is formed from the residue vibrational energy of the c-type 4-D pk internal to the corelets and 4-D core but that transport does not interfere with the final distribution of the x-type 4-D pk seeds. Indeed, radiation and matter could form within a void before the end of creation.

Einstein derived the 4-D geometry of his spacetime general relativity so that mass-energy influenced the curvature of spacetime and then that curvature interacted back on the dynamics of mass-energy, but there was no physical understanding of the machinery whereby mass-energy curved space and only the postulate that mass-energy followed a geodesic in that curved spacetime. Space could just expand without limit.

Here a different picture is postulated where, with a source term, 3-D space must obey a conservation law and Hubble's law must be derivable from the spatially generated expansion in which galaxies tend to participate. But there is more. If 3-D space is a medium that is being produced at all pk sites in that space and its outward flux is opposed by an object such as one responding to the gravity around a clump of mass, then there should be a counter expansion force proportional to the mass of the object and to its peculiar velocity. Such a gravity-limiting expansion force would have profound effects on the evolution of structure in the 3-D universe.

A tall order has been outlined for the expansion model and the internal physics of our 3-D universe, not the least of which is a new cosmic time that is asymmetric and provides a smooth transition with the creation model.

**Expansion Model**. Those equations of the SC-expansion model necessary to derive the predicted magnitude of supernova Ia were presented in Paper 1.[3] A few more equations will be added here and the connection to the 4-D planckton production will be made more explicit. Units of cgs are assumed unless otherwise stated.

Previously, we started with a universal constant,
$$\kappa = Gt^2\rho_T = Gt_0^2\rho_{T0} = 3/32\pi,^7 \tag{1}$$
that factored out of the derivation for the asymmetric time,
$$t = + (t_0^2\rho_{T0}/\rho_T)^{1/2}. \tag{2}$$
The zero subscripts designate present values and from the time derivatives of the scaling factors,
$$\rho_{T1} = \rho_r + \rho_m + \rho_x = \rho_T, \tag{3}$$
$$\rho_{T2} = 2\rho_r + 3/2\,\rho_m + \rho_x, \tag{4}$$
$$\rho_{T3} = 4\rho_r + 9/4\,\rho_m + \rho_x, \tag{5}$$
and
$$tH = (\rho_{T1}/\rho_{T2}). \tag{6}$$

From these expressions, the *deceleration* of the universe is,
$$q = -1 + 3[1 - 2/3(\rho_{T1}\rho_{T3}/\rho_{T2}^2)]R/((dR/dt)t). \tag{7}$$
By introducing the speed of light on both sides of an equation, the expansion rate is,

$$(dR/dt)/C = (R/Ct)(\rho_{T1}/\rho_{T2}). \tag{8}$$

To reduce this model from 5 adjustable parameters to only one, the following parameters were fixed: the density $\rho_{r0}=9.4\times10^{-34}$ was set by the temperature of the CBR, the density $\rho_{m0}=2.72\times10^{-31}$ was set by past nucleosynthesis calculations, the universal constant was set to $\kappa=3/32\pi$,[7] and the limiting value $((dR/dt)/C)_\infty$ was set to unity. The remaining parameters, including the present radius $R_0$ of the 4-D core, the present average value of the dark mass $\rho_{x0}$, the deceleration $q_0$, and the Hubble constant $H_0$, could all be calculated from the one input value for the present age $t_0$. An added input of $R/R_0=(1/(1+Z))$ gives the values of all parameters at any other $Z(t)$.

The correspondence with the production of 4-D pk has not been lost. The volume of the 4-D core is $V_{4U}=1/2\ \pi^2 R^4$ and the volume of its 3-D surface is $V_{3U}=2\pi^2 R^3$. The volume of a 4-D pk is $l_p^4$ and that of a 3-D pk is $l_p^3$, so the total number of pk in each volume is:

$$N_3=2\pi^2(R/l_p)^3 \text{ and } N_4=1/2\ \pi^2(R/l_p)^4, \tag{9}$$

where $l_p$ is the Planck length. From the derivatives with respect to cosmic time,

$$(dN_3/dt)/N_3 = 3H \text{ and } (dN_4/dt) = 4H\cdot N_4 = (m_3 C^2/h)((dR/dt)/C)(\rho_P/\rho_T), \tag{10}$$

For present parametric time t', $dN_4/dt'=m_3C^2/h$, then $(dR/dt)/C\sim1$, $\rho_P/\rho_T\sim10^{123}$. From these facts it was concluded [2] that non-columnar, vacuum spatial condensation cannot be measured.

## PREDICTIONS

**Present Parameter Values.** For an input age of $t_0$=13.5 Gy, the predicted reasonable present values for the cosmological parameters are: $R_0$=4388 Mpc; $q_0$=0.00842; $\Omega_m$ = 0.03075, $\Omega_{xo}$ = 0.2477; $\Omega_{T0}$ =0.2786; $H_0$=68.61 km s$^{-1}$ Mpc$^{-1}$, $(tH)_0$ =0.947; $(Ct/R)_0$=0.943 and $((dR/dt)/C)_0$=1.005. These values are presented to show, not only that our universe has nearly reached its steady-state expansion rate, but that dark mass has already become so dominant $\Omega_{xo}/\Omega_{T0}$=89 % that $(\rho_{T1}/\rho_{T2})\sim1$ and $(dR/dt)/C\sim1$.

Thus an amazing prediction of Eq. (8) is that in the fourth spatial dimension, approximately R=Ct. This is precisely the prescription that allows Einstein's 4-D geometry to be an excellent approximation locally and at the present time. This would not be so in the early universe because, as previously noted, at the end of creation $((dR/dt)/C)_{eoc} \sim 10^{24}$. Thus, a key feature of general relativity theory remains true in spite of many conceptual differences.

The computer program includes both the creation model and expansion model. The effects of variations of the one creation input parameter are presented elsewhere.[1,2] That input sets the time, temperature, densities, etc. at the end of creation and beginning of expansion, but otherwise, has no effect on the predictions of the expansion model. As shown in Fig. 2-2, an early computer run was made to sweep the entire range of size from the first 4-D pk to that or our universe when its radius is one thousand times its present size – 64 decades of $R/R_0$ from log $R/R_0$ = -61 to +3. With $t_0$=15.0 Gy, the run was made to show the production of both the total number $N_4$ of 4-D pk produced and the dark mass number $N_x$ x-type 4-D pk, as well as cosmic time.

One may wonder about the meaning of using the radius over the first one-half of the abscissa during creation but before our universe was even born. The "free" 4-D pk are considered incompressible, so given $N_4$, the radius of their total equivalent spherical volume is given by Eq. (9).

On the scale of Fig. 2-2, all of the detail of the last act of creation is lost between the points of log $R/R_0$ of $-27$ to $-26$, so to show that the curves are smoothly continuous, the narrow size range of log $R/R_0$ from $-26.24$ to $-26.19$ is shown by the small overlay graph. In Fig. 2-2 the beginning of creation is at $Z=10^{61}$ and the beginning of expansion is at $Z=10^{26}$. Many more such graphs of "early predictions" have been presented elsewhere.[1,2]

**Source of Gravity.** Our position of observation is still in epi-space, but now we are going to concentrate on the 4-D core to see what new physics is generated on its surface inside our 3-D universe as sketched in Fig. 2-3.

The new source of gravity is sketched in Fig. 2-4. The impact of the arriving m-D pk from epi-space produces an epi-force F toward the center of the 4-D core. For a large mass M that force produces a large dimple in the 4-D core. In turn for a smaller probe mass m positioned in the large dimple at distance r from M, the curvature of the large dimple at r produces a component 3-D force, f, towards M.

Figure 2-4a shows that $f=F\sin\theta$. Newton's theory of gravitation provides the value of f at any r but how do we separate the product $F\sin\theta$? Figure 2-4b suggests that $\sin\theta=1$ at the event horizon of a non-rotating black hole. So knowledge of mass M and the Schwarzschild radius $R_s$ of any non-rotating black hole gives the value of the epi-force per unit mass, $F/m$.

This epi-acceleration, due to the arriving m-D pk from epi-space, $a=F/m=-(\xi/4)(1/N_m)$ where $N_m=M/m_p$ is the large mass in units of the Planck mass $m_p$ and $\xi$ is a constant acceleration $\xi=(C^2/l_p)=5.569\times10^{53}$ cm s$^{-2}$ where C is the speed of light and $l_p$ is the Planck length. The second curvature factor $\sin\theta=\chi M^2/r^2$ where $\chi$ is a physical constant $\chi=4(l_p/m_p)^2=2.204\times10^{-56}$ cm$^2$g$^{-2}$ where $m_p$ is the Planck mass. Thus we have re-expressed Newton's equation in terms of the local curvature $\sin\theta$ of the 3-D universe. Note that although the force is Newtonian, geodesic motion on the curved 3-D surface of the 4-D core will not be Newtonian.

Feynman wrote that many have tried but no one has given any "machinery" behind gravity [6]. Now spatial condensation has provided "machinery" that must be tested and has indicated that the "attractive" characteristic of the force is only apparent.

At a 3-D radius outside the event horizon of a black hole, one can now express the curvature of our 3-D universe in terms of 4-D coordinates $(X_4,Y_4)$: $X_4=\int \cos\theta \, dr$ from $R_s$ to r and $Y_4=\int \sin\theta \, dr$ from $R_s$ to r.

**Black Hole Geometry.** Although the author has not seen the expression in the literature, the Hawking-Bekenstein entropy of a black hole can be written <u>exactly</u> as $S=(3k/8)(\rho_p/\langle\rho\rangle)$ where $\langle\rho\rangle$ is the average mass density within the event horizon and $\rho_p$ is the Planck density.[2] This expression suggests that the Planck density is a maximum and it is the cumulative force of the overburden at the mass M center that produces a black hole by penetration of the 4-D core.

With a few other simple assumptions about the distribution of mass within the 3-D lining inside a black hole, the 4-D geometry of the <u>inside</u> bottom of a black hole can now be derived[2]. With all distances expressed in units of the Schwarzschild radius, $R_s$, that 4-D geometry is shown for all non-rotating black holes in Fig.2-5.

The use of ordinary physics inside the black hole and the flat bottom in Fig.2-5 indicates that the 3-D gravitational acceleration f/m does go to zero at the center of a black hole in agreement with Newton's iron sphere theorem and Birkhoff's theorem[8]. That f/m does go to zero is demonstrated in Fig.2-6, which also confirms that F/m would decrease to the finite value of –ξ/4 for one Planck mass at the center.

So how does the curvature of 3-D space predicted by the SC-model compare to the curvature of the fictitious "hyperspace" of the general theory of relativity? A calculation by Kip Thorne in his recent book[9] makes that comparison possible. For a story about observers in five circular orbits of different circumferences about a black hole of mass M = 10 $M_{sun}$, Thorne calculated the "stretching factor" of the difference in gravity g across their bodies if they were aligned towards the center of the black hole. Assuming a height of 6 feet, the author converted these values to dg/dr to compare to the SC-calculated values of dg/dr for the curves of Fig.2-5. The comparison in Fig.2-7 shows good agreement. The SC-model also shows reasonable finite values of dg/dr inside the event horizon. General-relativistic "hyperspace" must predict a singularity.

The author thanks his good friend, Emeritus Professor Robert A. Piccirelli, for extensive discussions of the new physical concepts.

### NEXT PAPER 3

New physical concepts have been introduced in this spatial-condensation model and they will have a profound effect on our understanding of the evolution of large-scale structure in our universe. Those effects will be the subject of the next paper. In particular to be studied are the new 4-D reproducing (in place) dark mass with its growing gravitational force and the new cellular 3-D space that is being produced and must in turn generate an expansion force on any massive clump of matter with a substantial peculiar velocity. Many other new falsifiable predictions will be presented.

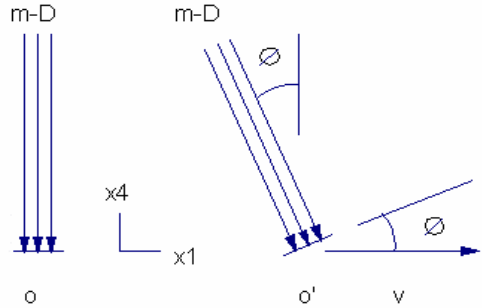

Fig. 2-1. Relative to radius R of the 4-D core, clock O is at rest in the comoving frame and clock O' (running slower) has additional kinetic energy with its persistent columns of m-D pk arriving at angle $\varphi = \cos^{-1}(1-(v/C)2)^{1/2}$.

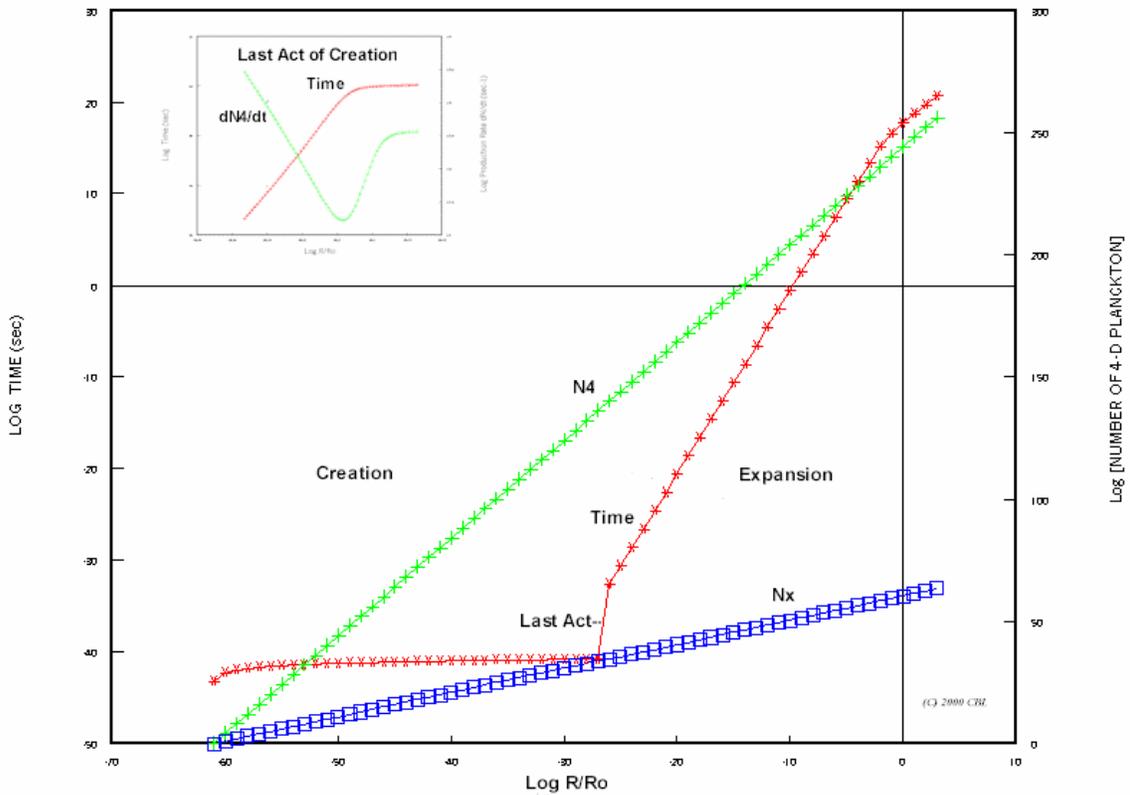

Fig. 2-2. Combining creation with expansion of our 3-D universe shows the smooth production of both c-type core-acceptable planckton $N_4$ and x-type 4-D pk $N_x$ of dark mass. With 64 decades of size on the abscissa, the rapid rise of time at log $R/R_0$ from –27 to –26 misses all of the detail, partly shown in the overlay for log $R/R_0$ from –26.24 to –26.19. Creation begins at $Z=10^{61}$, expansion at $Z=10^{26}$, and the present at $Z=0$.

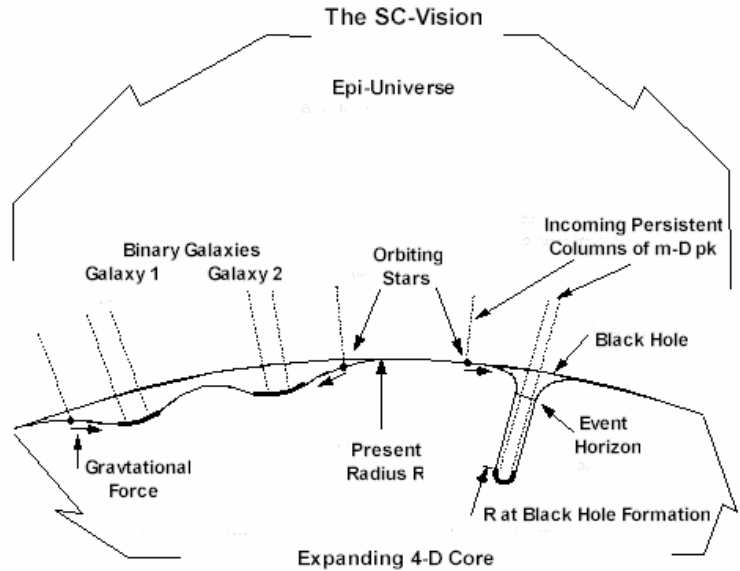

Fig.2-3. The vision of a new source of gravity is one of "pushing" towards a mass in contrast to the historic notion of "attraction." The incoming m-D pk impart a fraction of their 4-D momentum to their massive condensation site depending upon the local curvature of our 3-D space. A black hole is a dimple that has penetrated the 4-D core and is followed by the epi-space columns of m-D pk to the high-density mass at its base.

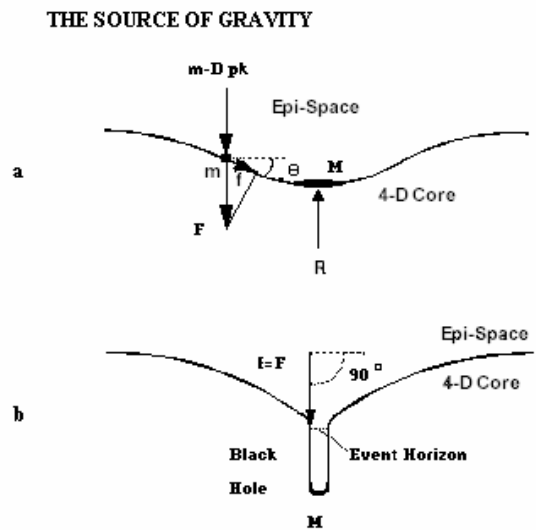

Fig.2-4. (a) Any mass M dimples the 4-D core so that arriving m-D pk to mass m creates a "gravitational force" toward M proportional to sin Ø. (b) A black hole is the ultimate dimple with ever increasing depth at rate (dR/dt)/C>1.

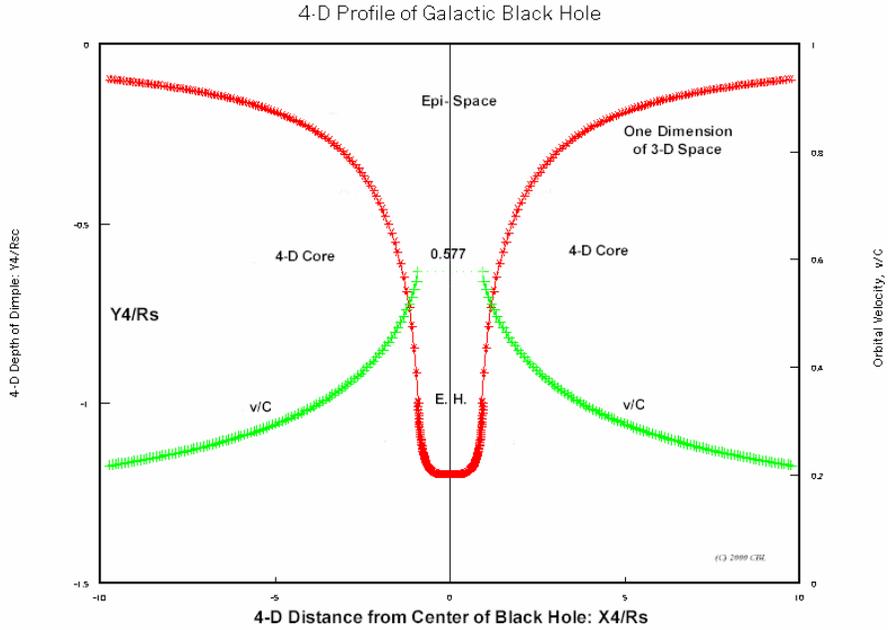

Fig.2-5. This four-dimensional (4-D) spatial profile of the inside of a black hole (<u>no singularity</u>) is one of the most significant predictions of the new *Spatial Condensation (SC)* model. The curvature of our 3-D space (only 1-D shown) outside the black hole, predicts the same acceleration toward the event horizon (E.H.) as the current model.

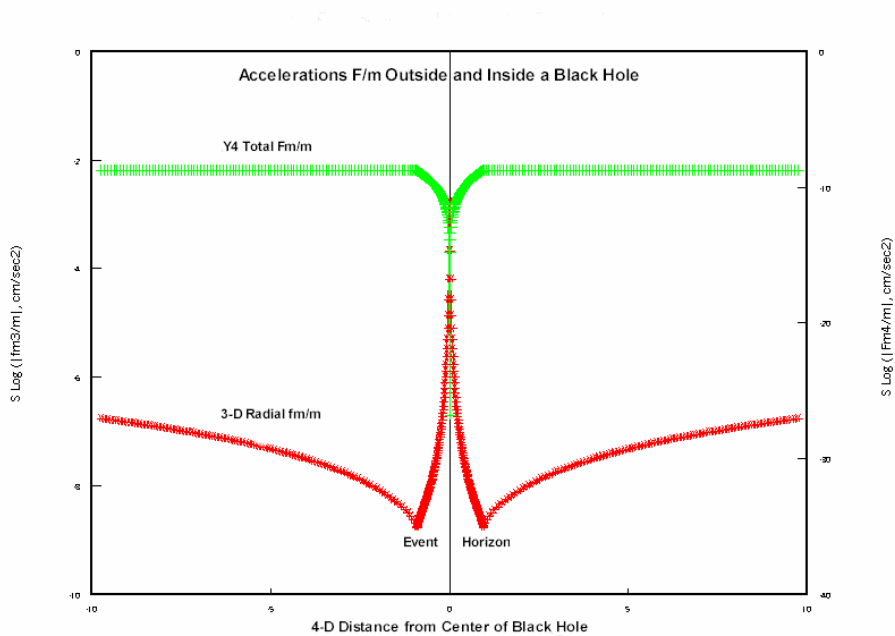

Fig.2-6. Inside an SC-black hole the impacting m-D force $F_{m4}/m$ toward the center of the universe increases in magnitude as r goes to zero. However the magnitude of the 3-D force $f_{m3}/m$ decreases to zero as r decreases to zero in agreement with the flat bottom of Fig. 2-5.

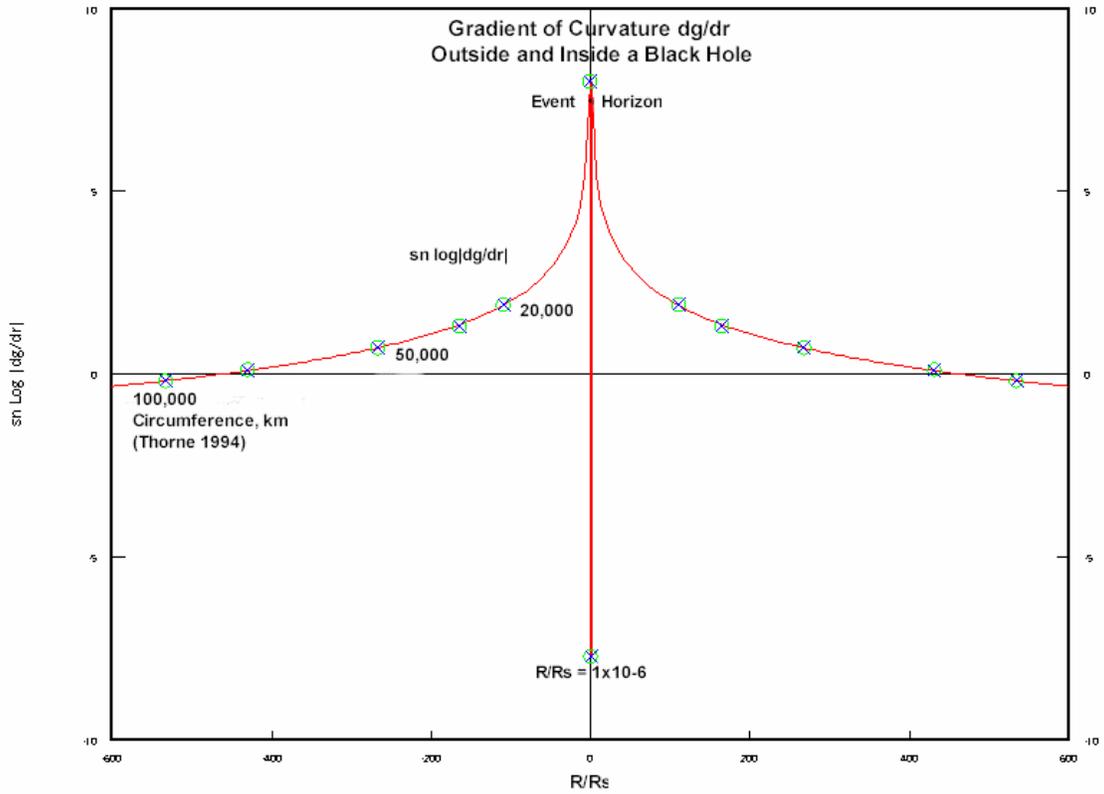

Fig.2-7. In agreement with the 4-D curvature of Fig.2-5 leading away from the event horizon of a black hole, the slope dg/dr (=d($f_m$/m)/dr) of the negative acceleration $f_m$/m is positive and decreasing in magnitude. At the event horizon for this black hole of mass M=10$M_{sun}$, dg/dr is maximum log(dg/dr) = +8.0 but inside, dg/dr abruptly changes to a negative constant, sn log |dg/dr| = -7.7 (sn=sign of argument). Agreement for dg/dr (outside B.H.) is shown compared with relativistic calculations of Thorne (1994), $R_s$=29.692 km).